\documentclass[11pt]{article}
\usepackage{latexsym}  
\usepackage{amssymb}
\usepackage{graphicx}
\usepackage{amsmath}

\begin{document}

\hspace*{9cm} {OU-HET-740/2012; MISC-2012-04}

\begin{center}
{\Large\bf Yukawaon Model with U(3)$\times$S$_3$ Family Symmetries}

\vspace{4mm}
{\bf Yoshio Koide$^a$ and Hiroyuki Nishiura$^b$}

${}^a$ {\it Department of Physics, Osaka University, 
Toyonaka, Osaka 560-0043, Japan} \\
{\it E-mail address: koide@het.phys.sci.osaka-u.ac.jp}

${}^b$ {\it Faculty of Information Science and Technology, 
Osaka Institute of Technology, 
Hirakata, Osaka 573-0196, Japan}\\
{\it E-mail address: nishiura@is.oit.ac.jp}

\date{\today}
\end{center}

\vspace{3mm}
\begin{abstract}
A new yukawaon model is investigated under a family symmetry
U(3)$\times$S$_3$.
In this model, all vacuum expectation values (VEVs)
of the yukawaons, $\langle Y_f\rangle$, are described in terms of 
a fundamental VEV matrix $\langle \Phi_0 \rangle$ as in the 
previous yukawaon model, but the assignments of quantum number for fields  
are different from the previous ones: the fundamental 
yukawaon $\Phi_0$ is assigned to $(3,3)$ of U(3)$\times$U(3), 
which is broken into $(3,1+2)$ of U(3)$\times$S$_3$, although quarks and 
leptons are still assigned to triplets of U(3) and yukawaons 
$Y_f$ are assigned to ${\bf 6}^*$ of U(3). 
Then, VEV relations among Yukawaons become more concise considerably 
than the previous yukawaon models.
By adjusting parameters, we can fit not only quark 
mixing parameters but also lepton mixing parameters 
together with their mass ratios.
\end{abstract}

PCAC numbers:  
  11.30.Hv, 
  12.15.Ff, 
  12.60.-i, 
\vspace{3mm}

\noindent{\large\bf 1 \ Introduction}

It is interesting to consider that the observed quark and lepton 
mass spectra and mixings can be understood from a ``family symmetry".
In the standard model (SM) of quarks and leptons, their mass spectra 
and mixings originate in the structures of the Yukawa coupling
constants.
If we intend to understand the observed mass spectra and mixings  
by adopting a non-Abelian gauge symmetry as a ``family symmetry", 
the symmetry will be explicitly broken at the
beginning because the Yukawa coupling constants have family indices. 
Therefore, most family symmetry models are based on a discrete 
symmetry.  
However, there is an easy way to consider non-Abelian gauge symmetry 
as the family symmetry: the Yukawa coupling constants are effective
ones $Y_f^{eff}$ ($f=u,d,e,\cdots$) and those are given by vacuum 
expectation values (VEVs) of new scalars $Y_f$:
$$
(Y_f^{eff})_{ij} = \frac{y_f}{\Lambda} \langle (Y_f)_{ij} \rangle .
\eqno(1.1)
$$
The fields $Y_f$ are called as ``yukawaon" \cite{yukawaon09}. 
The yukawaon model is a kind of ``flavon model" \cite{flavon}, 
but, differently from the conventional flavon model, 
the effective Yukawa coupling constant 
$Y_f^{eff}$ is essentially given by a $3\times 3$ VEV matrix 
of a single scalar $Y_f$. Here we do not consider that 
$Y_f^{eff}$ is given by a linear combination of scalars $Y_{A}$, 
$Y_{B}$, $\cdots$, which belong to different representations of
a family symmetry G, as $Y_f^{eff}= c_1 \langle Y_A \rangle + 
c_2 \langle Y_B \rangle +\cdots$. 
   
In the previous yukawaon model, the most characteristic feature is    
that all VEVs $\langle Y_f \rangle$ are  described in terms of 
only one fundamental VEV matrix $\langle \Phi_e \rangle$. 
For example, mass matrices for quarks and charged leptons are given as follows
 \cite{O3_09PLB}:
$$
\begin{array}{l}
M_e \propto \langle Y_e \rangle \propto \langle \Phi_e \rangle
\langle \Phi_e \rangle , \\
M_d \propto \langle Y_d \rangle \propto \langle \Phi_e \rangle
({\bf 1}+ a_d X) \langle \Phi_e \rangle , \\
M_u^{1/2}  \propto \langle \Phi_u \rangle \propto \langle \Phi_e \rangle
({\bf 1}+ a_u X) \langle \Phi_e \rangle ,
\end{array}
\eqno(1.2)
$$
where $\langle \Phi_e \rangle \propto  {\rm diag}(\sqrt{m_e},
\sqrt{m_\mu}, \sqrt{m_\tau})$ and  
$$
 {\bf 1} =  \left(
\begin{array}{ccc}
1 & 0 & 0 \\
0 & 1 & 0 \\
0 & 0 & 1
\end{array} \right) ,
\ \ \ 
 X \equiv \frac{1}{3}  \left(
\begin{array}{ccc}
1 & 1 & 1 \\
1 & 1 & 1 \\
1 & 1 & 1
\end{array} \right) .
\eqno(1.3)
$$
For neutrino sector, we consider a seesaw type of neutrino
mass generation $M_\nu = m_D M_R^{-1} m_D^T$, where 
the Dirac neutrino mass matrix $m_D$ and the right-handed 
neutrino Majorana mass matrix $M_R$ are given, respectively, by
$m_D \propto M_e$ and 
$$
M_R \propto \langle Y_R \rangle \propto
\langle Y_e \rangle \langle P_u \rangle \langle \Phi_u \rangle
+  \langle \Phi_u \rangle \langle P_u \rangle \langle Y_e \rangle
+ \cdots ,
\eqno(1.4)
$$
where $\langle P_u \rangle = {\rm diag}(+1,-1,+1)$ in the 
diagonal basis of the up-quark mass matrix $M_u$.
The model \cite{O3_09PLB} has roughly well described masses and 
mixings of the quarks and the leptons with a few parameters.
However, the fitting of the quark mixing has been somewhat unsatisfactory
compared with those of the neutrino mixing.
(For more precise parameter fitting, see Ref.\cite{HN-YK_11PRD}.)

In the present paper, we propose a new yukawaon model
where all yukawaon VEVs are described in terms of 
a new fundamental VEV matrix $\langle \Phi_0 \rangle$
similar to the previous model.
However, in contrast to the previous model (1.2), 
the charged lepton mass matrix $M_e$ is given by 
the same structure as those in the down- and up-quark sectors:
$$
\begin{array}{l}
M_e \propto \langle {Y}_e \rangle \propto 
\langle {\Phi}_0 \rangle ({\bf 1} + a_e X ) 
\langle {\Phi}_0 \rangle , \\
M_d \propto \langle {Y}_d \rangle \propto 
\langle {\Phi}_0 \rangle ({\bf 1} + a_d  X ) 
\langle {\Phi}_0 \rangle , \\
M_u^{1/2} \propto \langle \hat{Y}_u \rangle \propto 
\langle {\Phi}_0 \rangle ({\bf 1} + a_u X ) 
\langle {\Phi}_0 \rangle  . 
\end{array}
\eqno(1.5)
$$
[For more details, see Eqs.(4.9) - (4.14).]  
In the past yukawaon models, it has been assumed that 
the factor $({\bf 1}+ a_q X)$  ($q=u,d$) in Eq.(1.2)  
originates in VEV structures of additional fields. However,   
in the present paper, we assign the fundamental yukawaon $\Phi_0$ to  
$({\bf 1}+ {\bf 2})$ of a permutation symmetry S$_3$ \cite{S3}, 
and we consider that the factor $({\bf 1}+ a_q X)$ is due to 
a coefficient of $({\bf 1}+{\bf 2})\times ({\bf 1}+{\bf 2}) 
\rightarrow {\bf 1}$ under S$_3$. 
Therefore, it can naturally be understood that all the factors 
which are sandwiched by $\langle \Phi_0 \rangle$ are 
given by $({\bf 1}+a_f X)$ as shown in Eq.(1.5), instead of 
Eq.(1.2).  
The details will be discussed in the next section.

In Sec.3, we investigate possible superpotential forms for 
yukawaons, and in Sec.4, we summarize VEV relations in the 
present model. 
In Sec.5, we give parameter fittings for the observed 
lepton mixing (PMNS mixing \cite{PMNS}) and the observed quark 
mixing (CKM mixing \cite{CKM}) together with their mass ratios.
In the present model, we have 8 adjustable parameters 
$a_e$, $a_u e^{i\alpha_u}$, $a_d$, 
$\xi_\nu$, $m^0_d$ and $(\phi_1,\phi_2)$ for
16 observables (2 up-quark mass ratios, 2 down-quark
mass ratios, 2 neutrino mass ratios, 4 CKM mixing 
matrix parameters, 4+2 PMNS matrix parameters). 
The final section 6 is devoted to summary and discussions.

\vspace{2mm}

\noindent{\large\bf 2 \ Fundamental Yukawaon $\Phi_0$ and 
a permutation symmetry S$_3$}

First, we give a brief review of the factor $({\bf 1}+ a_f X)$ 
in S$_3$ symmetry.  
When we denote a doublet $(\psi_\pi, \psi_\eta)$ and 
a singlet $\psi_\sigma$ in a permutation symmetry S$_3$ \cite{S3}  
as
$$
\left( 
\begin{array}{c}
\psi_\pi \\
\psi_\eta 
\end{array} \right) = \left(
\begin{array}{c}
\frac{1}{\sqrt2} (\psi_1 -\psi_2) \\
\frac{1}{\sqrt6} (\psi_1 +\psi_2 -2\psi_3)
\end{array} \right) ,
\eqno(2.1)
$$
$$
\psi_\sigma =\frac{1}{\sqrt3}(\psi_1+\psi_2+\psi_3) ,
\eqno(2.2)
$$
the field $\psi = (\psi_1,\psi_2,\psi_3)$ is represented as
$$
\psi \equiv \left( 
\begin{array}{c}
\psi_1 \\
\psi_2 \\
\psi_3
\end{array} \right) = \left(
\begin{array}{ccc}
\frac{1}{\sqrt2} & \frac{1}{\sqrt6} & \frac{1}{\sqrt3} \\
-\frac{1}{\sqrt2} & \frac{1}{\sqrt6} & \frac{1}{\sqrt3} \\
0 & -\frac{2}{\sqrt6} & \frac{1}{\sqrt3} \\
\end{array} \right) \left(
\begin{array}{c}
\psi_\pi \\
\psi_\eta \\
\psi_\sigma
\end{array} \right)  .
\eqno(2.3)
$$
Then, a bilinear form $\psi\psi$ is invariant under the S$_3$ 
symmetry only when $\psi_a \xi_{ab} \psi_b$
is given by the form
$$
\psi_a \xi_{ab} \psi_b =
\psi_a ({\bf 1}+a_f X)_{ab} \psi_b ,
\eqno(2.4)
$$
where $a_f$ is a free parameter, and 
${\bf 1}$ and $X$ are defined by Eq.(1.3). 
The appearance of a free parameter $a_f$ is due to a reason that
there are two singlets which are composed of $\psi_a$, i.e. 
$\psi_\sigma \psi_\sigma$ and $(\psi_\pi,\psi_\eta)^T (\psi_\pi,\psi_\eta)$.

As seen from Eq.(2.4), we can assume that the fundamental yukawaon $\Phi_0$ is 
transformed as $({\bf 3}, {\bf 2}+{\bf 1})$ of U(3)$\times$S$_3$,
i.e. 
$$ 
\Phi^{0\, T}_{ia} ({\bf 1}+a_f X)^{ab} \Phi^0_{bj} .
\eqno(2.5)
$$
However, since the bilinear form (2.5) cannot have quantum numbers which
distinguish the sectors $f=u,d,e$ ($a_f$ are merely free parameters), 
we modify the expression (2.5) into 
$$ 
\Phi^{0\, T}_{i\alpha} S_f^{\alpha\beta} \Phi^0_{\beta j} ,
\eqno(2.6)
$$
where $S_f$ are fields and indices  $\alpha, \beta$ are of another U(3) 
symmetry (we denote it as U(3)$'$).
We assume that U(3)$'$ is broken into S$_3$ at an energy scale of $\Lambda'$ 
by non-vanishing VEV
$\langle S_f\rangle$ whose forms are given by
$$
 \langle S_f\rangle = v_{Sf} ({\bf 1}+a_f X) .
 \eqno(2.7)
$$
Here, we assume $\Lambda' \gg \Lambda$ (U(3) family symmetry is 
broken at $\Lambda$).  
Note that the indices $\alpha, \beta$ in the expression (2.6) are
of U(3)$'$, while the indices $a,b$ in the expression 
$$ 
(\Phi_0)^T_{ia} \langle S_f^{ab}\rangle (\Phi_0)_{bj} 
\eqno(2.8)
$$
are of S$_3$.
It is worthwhile to notice that in the conventional S$_3$ family model
quarks and leptons are assigned to (singlet+ doublet)'s of S$_3$, while
in the present model quarks and leptons are assigned to triplets of
the U(3) family symmetry, and only the fundamental yukawaon $\Phi_0$ 
is assigned to $({\bf 3}, {\bf 1}+{\bf 2})$ of U(3)$\times$S$_3$.

\vspace{2mm}

\noindent{\large\bf 3 \ Superpotential}

In the present model, would-be Yukawa interactions are given
as follows:
$$
W_Y = \frac{y_e}{\Lambda} e^c_i Y_e^{ij} \ell_j H_d 
+ \frac{y_\nu}{\Lambda} \nu^c_i Y_e^{ij} \ell_j H_u
+ \lambda_R \nu^c_i Y_R^{ij} \nu_j^c 
+ \frac{y_d}{\Lambda} d^c_i Y_d^{ij} q_j H_d  
+ \frac{y_u}{\Lambda} u^{c}_i Y_u^{ij} q_j H_u  ,
\eqno(3.1)
$$
where $\ell=(\nu_L, e_L)$ and $q=(u_L, d_L)$ are SU(2)$_L$ doublets.  
Under this definition of $Y_f$, the CKM mixing matrix and 
the PMNS mixing matrix are given by $V_{CKM}=U_u^\dagger U_d$ 
and $U_{PMNS}=U_e^\dagger U_\nu$, respectively, where
$U_f$ are defined by $U_f^\dagger M_f^\dagger M_f U_f
= D_f^2$ ($D_f$ are diagonal).  
In order to distinguish each yukawaon from others, we assume that 
$Y_f$ have different $R$ charges from each other together with $R$ charge conservation.
(Of course, the $R$ charge conservation is broken
at the energy scale $\Lambda'$.)
Here, we have assumed that the $R$ charge of the Dirac neutrino 
yukawaon $Y_\nu$ is identical with that of the charged lepton 
yukawaon $Y_e$, so that $Y_\nu$ is replaced with $Y_e$ for an 
efficient use of fields.

We assume the following superpotential for yukawaons:  
$$
W_e = \left[ \mu_e Y_e^{ij} +\frac{\lambda_e}{\Lambda} \bar{P}^{ik}
\hat{Y}^e_{kl} \bar{P}^{lj} \right] \Theta^e_{ji} 
+\left[ \mu'_e \hat{Y}^e_{ij} +\frac{\lambda'_e}{\Lambda} 
(\Phi_0)^T_{ia}  S_e^{ab} (\Phi_0)_{bj}  \right] 
\bar{\Theta}_e^{ji} ,
\eqno(3.2)
$$
$$
W_d = \left[ \mu_d Y_d^{ij} +\frac{\lambda_d}{\Lambda} \bar{P}^{ik}
\hat{Y}^d_{kl} \bar{P}^{lj} \right] \Theta^d_{ji} 
+\left[ \mu'_d \hat{Y}^d_{ij} +\frac{\lambda'_d}{\Lambda}
(\Phi_0)^T_{ia}  S_d^{ab} (\Phi_0)_{bj}
 + m_d^0 E_{ij} \right] \bar{\Theta}_d^{ji} ,
\eqno(3.3)
$$ 
$$
W_u =\frac{1}{\Lambda}  \left[ \lambda_u E_{ik} Y_u^{kl} E_{lj}
+ \lambda'_u \hat{Y}^u_{ik} E^{kl} \hat{Y}^u_{lj} \right] \bar{\Theta}_u^{ji} 
+ \left[ \mu'_u \hat{Y}^u_{ij}   
+\frac{\lambda^{\prime\prime}_u}{\Lambda}
(\Phi_0)^T_{ia}  S_e^{ab} (\Phi_0)_{bj}
 \right] \bar{\Theta}_u^{\prime\, ji} ,
\eqno(3.4)
$$
$$
W_R =\frac{1}{\Lambda} \left\{ \lambda_R E_{ik} Y_R^{kl} E_{lj} 
+ \lambda'_R  \left[
\hat{Y}^u_{ik} \bar{E}^{kl} \hat{Y}^e_{lj} +
 \hat{Y}^e_{ik} \bar{E}^{kl} \hat{Y}^u_{kj}
+\xi_\nu \left( {\rm Tr}[\hat{Y}_u \bar{E}] \hat{Y}^e_{ij}
+ {\rm Tr}[\bar{E} \hat{Y}^e] \hat{Y}^u_{ij} \right) 
\right] \right\} \bar{\Theta}_R^{ji} .
\eqno(3.5)
$$
(In this paper, we use $\hat{Y}^u$ instead of $\Phi_u$ unlike the
previous papers, since we treat the up- and down-quark sectors in the same way  
$\hat{Y}^u \leftrightarrow \hat{Y}^d$.)  
Here we have assumed that the $R$ charges satisfy the following relation
$$
R(Y_e) -R(Y_d) = R(\hat{Y}^e) - R(\hat{Y}^d) =R(S_e) -R(S_d) .
\eqno(3.6)
$$
The VEVs of the introduced fields $E$, $\bar{E}$ and $\bar{P}$ are described by the
following superpotential 
$$
W_{E,P} = \frac{\lambda_1}{\Lambda} {\rm Tr}[\bar{E} E \bar{P} P] 
+\frac{\lambda_2}{\Lambda} {\rm Tr}[\bar{E} E]  {\rm Tr}[\bar{P} P] ,
\eqno(3.7)
$$
which leads to 
$$
\langle E \rangle \langle \bar{E} \rangle \propto {\bf 1} , \ \ \ \ 
\langle P \rangle \langle \bar{P} \rangle \propto {\bf 1} .
\eqno(3.8)
$$
We take specific solutions of Eq.(3.8):
$$
\frac{1}{v_E} \langle E \rangle = \frac{1}{\bar{v}_E} \langle \bar{E} \rangle 
= {\bf 1} , 
\eqno(3.9)
$$
$$
\frac{1}{v_P} \langle P \rangle = \frac{1}{\bar{v}_P^*} 
\langle \bar{P} \rangle^\dagger = {\rm diag}( e^{-i\phi_1}, e^{-i\phi_2}, 1),
\eqno(3.10)
$$
as the explicit forms of $\langle E \rangle$, $\langle \bar{E} \rangle$ 
and $\langle \bar{P} \rangle$. 
Here, we have assigned $R$ charges for those fields as
$$
R(\bar{E}) +R(E) +R(\bar{P}) +R(P) =2 .
\eqno(3.11)
$$
Since we consider $R(\bar{E}) +R(E) \neq R(\bar{P}) +R(P)$, 
terms ${\rm Tr}[\bar{E} E \bar{E} E]$,  
${\rm Tr}[\bar{P} P \bar{P} P]$, and so on are forbidden
by the $R$ charge conservation. 
However, the absence of the term  
${\rm Tr}[\bar{E} P]  {\rm Tr}[\bar{P} E]$ in Eq.(3.7)
must be assumed ad hoc.

In the superpotential $W_d$, Eq.(3.3), we have taken
$$
R(E) = R(\hat{Y}^d) , 
\eqno(3.12)
$$
and we have added the $m_0^d E_{ij}$ term to $\hat{Y}^d_{ij}$.
The relation (3.12) has been ad hoc assumed in order to adjust
the quark mass ratio $m_{d1}/m_{d2}$. 
Such the assignment is not contradict with whole $R$ charge 
assignments of the fields.
However, once the assignment (3.12) is done, then, we cannot add $E$ term
to $\hat{Y}^e$ and/or $\hat{Y}^u$ because of the $R$ charge conservation.

We list whole fields in the present model in Table 1.
As seen in Table 1, the sum of the anomaly coefficients $A$ is 
$\sum A = -9$.  
In order to be $\sum A=0$, we need further three fields,
$T^1_{ia}$, $T^2_{ia}$ and $T^3_{ia}$. 
For the time being, we do not specify the roles of these fields in this model.

\begin{table}
\begin{center}

\begin{tabular}{c|ccccccccc}\hline   
 & $\ell_i$ & $e^c_i$ & $\nu^c_i$ & $q_i$ & $u^c_i$ & $d^c_i$ &
$H_u$ & $H_d$ & $\Phi^0_{ia}$ \\ \hline
U(3) & ${\bf 3} $  & ${\bf 3} $  & ${\bf 3} $  & ${\bf 3} $  & ${\bf 3} $ 
 & ${\bf 3} $  & ${\bf 1} $  & ${\bf 1} $ & ${\bf 3}$ \\
U(3)$'$ & ${\bf 1}$ & ${\bf 1}$ & ${\bf 1}$ & ${\bf 1}$ &
 ${\bf 1}$ & ${\bf 1}$ & ${\bf 1}$ & ${\bf 1}$ & ${\bf 3}$ \\ \hline
 \end{tabular}
 
\begin{tabular}{cccccccccc}\hline 
$Y_e^{ij}$ & $\Theta^e_{ij}$ & $\hat{Y}^e_{ij}$ & $\bar{\Theta}_e^{ij}$ 
& $S_e^{ab}$ & 
$Y_d^{ij}$ & $\Theta^d_{ij}$ & $\hat{Y}^d_{ij}$ & $\bar{\Theta}_d^{ij}$ 
& $S_d^{ab}$  \\  \hline 
${\bf 6}^* $  & ${\bf 6} $  & ${\bf 6} $  & ${\bf 6}^*$  & ${\bf 1}$ 
 & ${\bf 6}^* $  & ${\bf 6} $  & ${\bf 6} $  & ${\bf 6}^*$  & ${\bf 1}$ \\
 ${\bf 1}$ & ${\bf 1}$ & ${\bf 1}$ & ${\bf 1}$ & ${\bf 6}^*$ & ${\bf 1}$ &
 ${\bf 1}$ & ${\bf 1}$ & ${\bf 1}$ & ${\bf 6}^*$ \\ \hline
\end{tabular}

\begin{tabular}{cccccccccccc}\hline 
$Y_u^{ij}$ & $\bar{\Theta}_u^{ij}$ & $\hat{Y}^u_{ij}$ & 
$\bar{\Theta}_u^{\prime\, ij}$ & $S_u^{ab}$ &
 $Y_R^{ij}$ &  $\bar{\Theta}_R^{ij}$ & 
$\bar{P}^{ij}$ & $P_{ij}$ & $E_{ij}$ & $\bar{E}^{ij}$ \\ \hline
${\bf 6}^* $  & ${\bf 6}^* $  & ${\bf 6} $  & ${\bf 6}^*$  & ${\bf 1}$ 
 & ${\bf 6}^* $  & ${\bf 6}^* $  & ${\bf 6}^* $  & ${\bf 6} $  &
 ${\bf 6} $  & ${\bf 6}^* $  \\
 ${\bf 1}$ & ${\bf 1}$ & ${\bf 1}$ & ${\bf 1}$ & ${\bf 6}^*$ & 
 ${\bf 1}$ & ${\bf 1}$ & ${\bf 1}$ & ${\bf 1}$ & ${\bf 1}$ & ${\bf 1}$ 
\\ \hline
\end{tabular}

\end{center}

\caption{
Quantum numbers of the fields 
}

\end{table}

\vspace{2mm}

\noindent{\large\bf 4 \ VEV relations}

We assume that our vacuum always takes 
$\langle \theta_A \rangle=0$
($A=e,u,d,\cdots$).
Therefore, VEV relations are obtained from the 
SUSY vacuum conditions $\partial W/\partial \Theta_A =0$.
Since relations from another SUSY vacuum conditions 
$\partial W/\partial Y_e =0$ and so on always 
include one of $\langle \theta_A \rangle$, 
such the relations do not have meaning except for Eq.(3.7).
Since we assume that the SUSY breaking is caused by gauge mediation 
(except for family gauge symmetries), we consider that 
our VEV relations in the yukawaon sector are satisfied 
until a low energy scale.

From the superpotential terms (3.2)-(3.5), we obtain
the following VEV relations among the yukawaons:
$$
\langle Y_e^{ij} \rangle = - \frac{\lambda_e}{\mu_e \Lambda}
\langle \bar{P}^{ik} \rangle \langle \hat{Y}_{kl}^e \rangle
\langle \bar{P}^{lj} \rangle ,
\eqno(4.1)
$$
$$
 \langle \hat{Y}^e_{ij} \rangle  =-\frac{\lambda'_e}{\mu'_e\Lambda} 
 \langle (\Phi_0)^T_{ia} \rangle 
 \langle S_e^{ab} \rangle  \langle (\Phi_0)_{bj} \rangle , 
\eqno(4.2)
$$
$$
\langle Y_d^{ij} \rangle = - \frac{\lambda_d}{\mu_d \Lambda}
\langle \bar{P}^{ik} \rangle \langle \hat{Y}_{kl}^d \rangle
\langle \bar{P}^{lj} \rangle ,
\eqno(4.3)
$$
$$
 \langle \hat{Y}^d_{ij} \rangle  =-\frac{\lambda'_d}{\mu'_d\Lambda}
\langle (\Phi_0)^T_{ia} \rangle 
 \langle S_d^{ab} \rangle  \langle (\Phi_0)_{bj} \rangle
 + \frac{m_d^0}{\mu'_d} \langle E_{ij} \rangle , 
\eqno(4.4)
$$
$$
 \langle Y_u^{ij} \rangle  = - 
\frac{ \lambda'_u}{\lambda_u} (\langle E^{-1} \rangle)^{ik}
\langle \hat{Y}^u_{kl} \rangle \langle E^{lm}\rangle 
\langle \hat{Y}^u_{mn} \rangle  (\langle E^{-1} \rangle)^{nj} ,
\eqno(4.5)
$$
$$
 \langle \hat{Y}^u_{ij} \rangle  =-\frac{\lambda'_u}{\mu'_u\Lambda} 
 \langle (\Phi_0)^T_{ia} \rangle 
 \langle S_u^{ab} \rangle  \langle (\Phi_0)_{bj} \rangle  , 
\eqno(4.6)
$$
$$
 \langle  Y_R^{ij} \rangle  =- \frac{\lambda'_R}{\lambda_R}  
 ( \langle E^{-1} \rangle)^{ik} \left[
 \langle \hat{Y}^u_{kl}\rangle  \langle \bar{E}^{lm}\rangle 
\langle  \hat{Y}^e_{mn}\rangle +
 \langle  \hat{Y}^e_{kl}\rangle  \langle \bar{E}^{lm}\rangle 
 \langle \hat{Y}^u_{mn}\rangle 
 \right.
 $$
 $$
 \left. 
+\xi_\nu  \left( {\rm Tr}[ \langle \hat{Y}_u \rangle \langle \bar{E}\rangle] 
 \langle \hat{Y}^e_{kn}\rangle
+ {\rm Tr}[ \langle \bar{E}\rangle \langle \hat{Y}^e\rangle]
 \langle \hat{Y}^u_{kn}\rangle \right) 
\right]  ( \langle E^{-1} \rangle)^{nj} ,
\eqno(4.7)
$$
where $\langle S_f \rangle$, $\langle E \rangle$, 
$\langle \bar{E}\rangle$ and $\langle \bar{P} \rangle$ are given by
Eqs.(2.7), (3.9) and (3.10), respectively.
We have assumed that the VEV matrix $\langle \Phi_0 \rangle$ is
diagonal in the basis in which the VEV matrix  $\langle S_f \rangle$
take the form $({\bf 1} + a_f X)$, i.e. 
$$
\langle \Phi_0 \rangle = {\rm diag}(v_1, v_2, v_3).
\eqno(4.8)
$$

In the present model, common coefficients are not important.
Therefore, when we omit those coefficients,  quark and lepton 
mass matrices are given as follows:
$$
M_e = \langle \bar{P}\rangle \hat{M}_e \langle \bar{P} \rangle ,
\eqno(4.9)
$$
$$
 \hat{M}_e  = \langle \Phi_0^{T} \rangle 
 \langle S_e \rangle  \langle \Phi_0 \rangle .
 \eqno(4.10)
$$
$$
M_d = \langle \bar{P} \rangle \left( \langle \Phi_0^{T} \rangle 
 \langle S_d \rangle  \langle \Phi_0 \rangle 
 + m^0_d {\bf 1} \right) \langle \bar{P} \rangle ,
\eqno(4.11)
$$
$$
M_u^{1/2} = \langle \Phi_0^{T} \rangle 
 \langle S_u \rangle  \langle \Phi_0 \rangle , 
\eqno(4.12)
$$
$$
M_\nu = M_e M_R^{-1} M_e^T ,
\eqno(4.13)
$$
$$
M_R = M_u^{1/2} \hat{M}_e + \hat{M_e} M_u^{1/2} +
+\xi_\nu  \left( {\rm Tr}[ M_u^{1/2} ] \hat{M}_e
+ {\rm Tr}[ \hat{M}^e] M_u^{1/2} \right)  .
\eqno(4.14)
$$
For numerical calculations in the next section, 
we will use  dimensionless expressions 
$\langle \Phi_0 \rangle = {\rm diag}(x_1, x_2, x_3)$ 
with $x_1^2+x_2^2+x_3^2=1$, 
$\langle \bar{P} \rangle = {\rm diag}(e^{i\phi_1}, 
e^{i\phi_2}, 1)$ and $\langle S_f \rangle = {\bf 1}
+a_f X$ in Eqs.(4.9) - (4.14).
(Therefore, $m^0_d$ in Eq.(4.11) is also a dimensionless
parameter.)

\vspace{2mm}

\noindent{\large\bf 5 \ Parameter fitting}

For simplicity, we assume that the parameter 
$a_d$ in the down-quark sector is real as well as 
$a_e$ in the charged lepton sector, so that only $a_u$ is complex.
For the convenience, hereafter, we denote $a_u$ anew 
as $a_u e^{i\alpha_u}$ ($a_u$ and $\alpha_u$ are
real). 
Then, in this model, we have 8 adjustable parameters 
$a_e$, $a_u e^{i\alpha_u}$, $a_d$, 
$\xi_\nu$, $m^0_d$ and $(\phi_1,\phi_2)$ for
16 observables [2 up-quark mass ratios, 2 down-quark
mass ratios, 2 neutrino mass ratios, 4 CKM mixing 
matrix parameters, 4+2 PMNS matrix parameters]. 
Since the observed charged lepton mass ratios are 
used as input values, we do not count them as adjustable 
parameters.  
The parameter $m^0_d$ is used only in order to 
fit the down-quark mass ratio $m_d/m_s$.
(In other words, other observables are insensitive
to the value of $m^0_d$.)
Therefore, 
if we do not count this parameter $m^0_d$, we have 7 
parameters for 15 observables. 

\vspace{2mm}

{\large\bf 5.1 \ PMNS mixing}  

In order to predict the neutrino mixing (PMNS mixing) parameters, 
we have 6 parameters [$a_e$, $\xi_\nu$, $a_u e^{i \alpha_u}$ and
$(\phi_1,\phi_2)$]. 
At present, we know only  5 observed quantities [2 up-quark mass ratios,
1 neutrino mass ratio $R_\nu = \Delta m^2_{21}/\Delta m^2_{32}$, 
3 neutrino mixing angles,  $\sin^2 2\theta_{atm}$, 
$\tan^2 \theta_{solar}$ and 
$\sin^2 2\theta_{13}$] among 10 observable quantities [2 up-quark mass ratios, 
2 neutrino mass ratios, 4+2 PMNS mixing parameters]. 
Therefore, in principle, we cannot determine the parameter 
values from the observed quantities. Instead, we use the observables which are 
described by parameters as few as possible.
To start with, let us use up-quark mass ratios $r^u_{12}= \sqrt{m_u/m_c}$
and $r^u_{23}= \sqrt{m_c/m_t}$ which are described by 
3 parameters $a_e$ and $a_u e^{i\alpha_u}$.
The observed values of $r^u_{12}$ and $r^u_{23}$ are
as follows \cite{q-mass}:
$$
r^u_{12} \equiv \sqrt{\frac{m_u}{m_c}} 
= 0.045^{+0.013}_{-0.010} , \ \ \ \ 
r^u_{23} \equiv \sqrt{\frac{m_c}{m_t}}
=0.06 \pm 0.005 ,
\eqno(5.1)
$$
at $\mu=m_Z$. 
We illustrate a relation $(a_u, \alpha_u)$ and $a_e$ which 
satisfy $r^u_{12}=0.045$ and $r^u_{23}=0.060$ in Fig.~1.
As seen in Fig.~1, the these parameters are bounded 
in $a_u = -(1.3 - 1.8)$, $a_u=0^\circ - 10^\circ$ and
$a_e = 0 - 10$.
(Since the observed center values  have large errors,
these bounds should not be taken rigidly.)

\begin{figure}[t!]
  \includegraphics[width=60mm,clip]{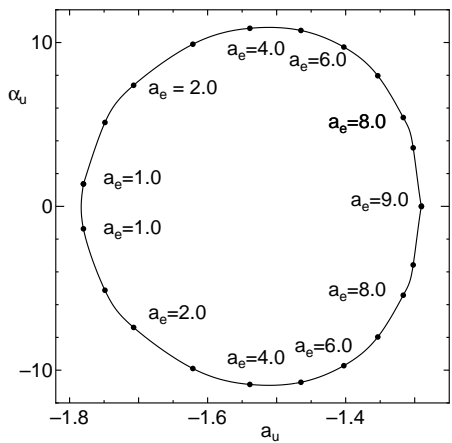}
  \caption{ Parameter values $(a_u, \alpha_u)$ and $a_e$
which satisfy the observed values of up-quark mass ratios
$\sqrt{m_u/m_c}=0.045$ and $\sqrt{m_c/m_t}=0.060$
}
  \label{fig1}
\end{figure}

Next, we focus on the neutrino mixing $\sin^2 2\theta_{atm}$. 
We find that it is insensitive to $\alpha_u$ and $\xi_\nu$, 
although it is described by parameters
$a_e$, $a_u e^{i\alpha_u}$, $\xi_\nu$ and $(\phi_1,\phi_2)$ in our model.
In Fig.~2, we show curves in the $(a_u, a_e)$ plane which satisfy  
$\sin^2 2\theta_{atm}=0.90$, $0.92$, $0.94$ and $0.96$, 
respectively. 
Here, for convenience, we have fixed $(\phi_1,\phi_2)$ as 
$(\phi_1,\phi_2)=(180^\circ, 180^\circ)$.
(For a case of $(\phi_1,\phi_2)=(0^\circ, 0^\circ)$, 
we could not find reasonable parameter solutions at all.) 
Here, we have used $\xi_\nu=0.0025$ and $\alpha_u = 8^\circ$
and $10^\circ$, and $(\phi_1, \phi_2)=(180^\circ, 180^\circ)$
tentatively. 
As seen in Fig.~2, two curves ($\alpha_u = 8^\circ$
and $10^\circ$) are almost degenerated. 
For reference, we also show 
a curve which satisfies reasonable up-quark mass ratios.
Since we cannot take $a_u > -1.3$ as we have shown in Fig.~1,
we cannot obtain a value more than $\sin^2 2\theta_{atm} = 0.97$.
If we want to obtain a value of $\sin^2 2\theta_{atm}$ as 
large as possible, we must take a parameter set 
$$
(a_e, a_u, \alpha_u) \sim (8, -1.35, \pm 6^\circ) .
\eqno(5.2)
$$
For this parameter set (5.2), we can obtain 
$\tan^2 \theta_{solar} \sim 0.5$ and $R \sim 0.04$.
More accurate predicted values will be given after
we discuss the CKM mixing values in the next subsection.

\begin{figure}[t!]
  \includegraphics[width=60mm,clip]{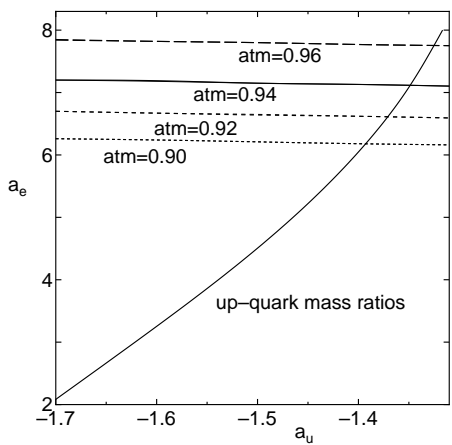}
  \caption{ Parameter values $(a_u, a_e)$ 
which satisfy $\sin^2 2\theta_{atm}=0.90$, $0.92$, 
$0.94$, $0.96$, respectively. (In the figure, ``atm" 
means $\sin^2 2\theta_{atm}$.)
For reference, the curve which satisfies 
up-quark mass ratios
$\sqrt{m_u/m_c}=0.045$ and $\sqrt{m_c/m_t}=0.060$
are also shown.
}
  \label{fig2}
\end{figure}
 
\vspace{2mm}

{\large\bf 5.2 \ CKM mixing}

The 4 observables in the CKM mixing matrix depend on 
3 parameters $a_d$ and $(\phi_1, \phi_2)$
in addition to the parameter values (5.2).
Although the parameter $a_d$ can be 
determined by two down-quark mass rations \cite{q-mass}
$$
r^d_{12} \equiv \frac{m_d}{m_s} = 0.053^{+0.005}_{-0.003} ,
\ \ \ 
r^d_{23} \equiv \frac{m_s}{m_b} = 0.019^{+0.006}_{-0.006} ,
\eqno(5.3)
$$
we cannot give reasonable fitting for the 
CKM mixing and the down-quark mass ratios simultaneously. 
Therefore, we fit only the value of $r^d_{23}$ by 
$a_d$, and the value $r^d_{12}$ is 
adjusted by the parameter $m^0_{d}$ which
does not affect the CKM mixing parameters. 

Although we have roughly obtained the parameter values of 
$(a_e, a_u, \alpha_u)$ in Eq.(5.2) from $r^u_{23}$, $r^u_{12}$, and $\sin^22\theta_{atm}$, 
the predicted values for 
CKM mixing are also dependent on the parameter values 
$(a_e, a_u, \alpha_u)$. We consider that the observed values for the quark 
mass ratios are still controversial. 
Therefore, we search a set of reasonable parameter values for 
$[a_e, (a_u, \alpha_u), \xi_\nu,  a_d, (\phi_1, \phi_2)]$ 
that can give the observed PMNS and CKM mixings within 
one sigma and quark mass ratios within two sigma, 
keeping the result (5.2) in mind. 
Namely, After searching a rough value of the parameter $a_d$ which can 
give reasonable CKM mixings and the down-quark mass ratio $r^d_{23}$, 
we do a fine tuning of parameter values for 
$[a_e, (a_u, \alpha_u), a_d, (\phi_1, \phi_2)]$ 
with help of figures demonstrated in Fig.~3. 

\begin{figure}[t!]
  \includegraphics[width=70mm,clip]{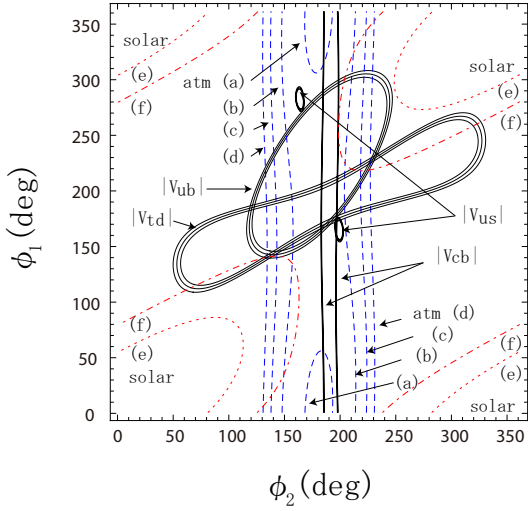}
  \caption{ Contour Plots in the ($\phi_1$, $\phi_2$) parameter 
plane, which are shown by using experimental constraints on
  $|V_{us}|=0.2252 \pm 0.0009$, $|V_{cb}|=0.0406\pm 0.0013$, $|V_{ub}|=0.00389 \pm 0.00044$, and  $|V_{td}|=0.0084 \pm 0.0006$ and by taking some values for atm 
$\equiv \sin^22\theta_{atm}$ and  $\equiv \tan^2\theta_{solar}$.
  (a): $\sin^22\theta_{atm}$=0.98, (b): $\sin^22\theta_{atm}$=0.96,  
(c): $\sin^22\theta_{atm}$=0.94,  
and  (d): $\sin^22\theta_{atm}$=0.92,  which are shown by dashed 
lines, and solar$\equiv \tan^2\theta_{solar}$ which is shown by dotted  
or dot-dashed line. (e): $\tan^2\theta_{solar}=0.495$ (dotted line), 
and (f): $\tan^2\theta_{solar}=0.515$ (dot-dashed line). 
Here we take the values for the parameter set of $[a_e, (a_u, \alpha_u), \xi_\nu, a_d, m^0_{d}]$ given in Eq.~(5.10). 
}
  \label{Contour_Plots}
\end{figure}

As a result, we obtain the following predictions:
$$
\sin^2 2\theta_{atm}=0.965, \ \ \ 
\tan^2 \theta_{solar}= 0.522, \ \ \
\sin^2 2\theta_{13} = 0.027 , 
\eqno(5.4)
$$
$$
\delta_{CP}^{\ell}= -177^\circ  \ \ \ (J^{\ell} = -9.3 \times 10^{-4} ), 
\eqno(5.5)
$$
$$
R_\nu =\frac{\Delta m^2_{solar}}{\Delta m^2_{atm}} = 0.044,
\eqno(5.6)
$$
$$
|V_{us}|=0.2240, \ \ \ |V_{cb}|=0.0404, \ \ \ 
|V_{ub}|=0.00409, \ \ \ |V_{td}|= 0.00823 ,
\eqno(5.7)
$$
$$
\delta_{CP}^{q}= 65.3^\circ   \ \ \ (J^{q} = 3.3 \times 10^{-5} ), 
\eqno(5.8)
$$
$$
r^u_{12} =0.0422, \ \ \ r^u_{23}= 0.0659 , \ \ \ 
r^d_{12} = 0.0530, \ \ \ r^d_{23}=0.0360,
\eqno(5.9)
$$
under the parameter values
$$
a_e=8.5, \ a_u=-1.32, \  \alpha_u=-6.5^\circ, \  a_d=17, \ 
m^0_d=0.0113, 
$$
$$
 \xi_\nu=0.0019, \  (\phi_1,\phi_2)=(174.5^\circ,195.9^\circ).
\eqno(5.10)
$$
Although the predicted value $\tan^2 \theta_{solar}= 0.522$ is 
somewhat large compared with the observed value \cite{PDG10}
$\tan^2 \theta_{solar}=0.468^{+0.048}_{-0.029}$, it is 
consistent with the KamLAND data \cite{KamLAND08} 
$\tan^2 \theta_{solar}=
0.56^{+0.10}_{-0.07}{\rm (stat)}^{+0.10}_{-0.06}{\rm (syst)}$.
The predicted value $\sin^2 2\theta_{13} = 0.027$ is small
compared with the T2K data \cite{T2K11} 
$0.03 < \sin^2 2\theta_{13} <0.28$
for $\delta_{CP}^{\ell}=0$, but it is still not ruled out because 
our model predicts  $\delta_{CP}^{\ell}= -177^\circ$. 
Also, the predicted value $r^d_{23}=0.0352$ is larger than the 
center value given in Eq.(5.3) by $3\sigma$. 
Rather, the predicted value is near to the old value 
$r^d_{23} = 0.031 \pm 0.005$ given in the second literature 
in Ref.\cite{q-mass}.  
We consider that the value of $m_s$ is still controversial.   

We can also predict neutrino masses
$$
m_{\nu 1} \simeq 0.0056\ {\rm eV}, \ \ m_{\nu 2} \simeq 0.0118 \ {\rm eV}, 
\ \ m_{\nu 3} \simeq 0.0507 \ {\rm eV}  ,
\eqno(5.11)
$$
by using the input value \cite{MINOS08}
$\Delta m^2_{32}\simeq 0.00243$ eV$^2$.
We also predict the effective Majorana neutrino mass \cite{Doi1981} $\langle m \rangle$ 
in the neutrinoless double beta decay as
$$
\langle m \rangle =\left|m_1 U_{e1}^2 +m_2 U_{e2}^2 
+m_3 U_{e3}^2\right| \simeq 3.2 \times 10^{-4}\ {\rm eV}.
\eqno(5.12)
$$
 
\vspace{2mm}

\noindent{\large\bf 6 \ Concluding remarks}

In conclusion, by assuming U(3)$\times$S$_3$ family symmetries,
we have proposed a new yukawaon model in which not only the quark
mass matrices $M_u^{1/2}$ and $M_d$ but also the charged lepton
mass matrix $M_e$ is given by a form 
$\langle \Phi_0 \rangle ({\bf 1}+a_f X) \langle \Phi_0 \rangle$ 
($\Phi_0$ is a fundamental yukawaon) as shown in Eqs.(4.9) - 
(4.14).   
The model have only 7 parameters for 15 observables. However, we 
have obtained reasonable predictions. 

Since all the mass matrices for quarks and charged leptons are given by the same form 
$\langle \Phi_0 \rangle ({\bf 1}+a_f X) \langle \Phi_0 \rangle$,
we can consider a possibility that if we define 
$\langle Y_0 (a_f) \rangle = \langle \Phi_0 \rangle 
({\bf 1}+a_f X) \langle \Phi_0 \rangle$ by introducing a yukawaon $Y_0$, all the effective Yukawa 
coupling constants $Y_f^{eff}$ can be given by a single yukawaon 
VEV $\langle Y_0 (a_f) \rangle$.
However, this idea runs into a stone wall, because we cannot 
distinguish $Y_0(a_d)$ from $Y_0(a_e)$ in the expression of 
$\langle Y_R \rangle$, Eq.(4.14). 
Besides, the VEV of the ``single" yukawaon $Y_0$ is dependent 
on the parameter $a_f$.
This is contradictory to the idea in the original yukawaon 
model that $Y_f^{eff}$ is given by a single yukawaon VEV 
without including a free parameter.
Therefore, in the present paper, we have considered that
the yukawaons $Y_f$ are distinguished by the fields $S_f$ in 
$\Phi_0 S_f \Phi_0$ as shown in Eq.(2.8).

One of motivations of the yukawaon model \cite{yukawaon09} was to 
understand a charged lepton mass relation \cite{K-relation}
$m_e +m_\mu +m_\tau = (2/3)(\sqrt{m_e} +\sqrt{m_\mu} 
+\sqrt{m_\tau})^2$ by considering 
$\langle Y_e \rangle \propto \langle \Phi_e \rangle 
\langle \Phi_e \rangle$.  
However, in the present model, since $\langle Y_e \rangle$
is given by $\langle \Phi_0 \rangle 
({\bf 1}+a_e X) \langle \Phi_0 \rangle$, 
the scenario for the charged lepton mass relation must 
be abandoned, and we have to search an alternative scenario
for the charged lepton mass relation.  
Nevertheless, apart from this problem, it seems that 
the present new yukawaon model offers us 
a promising hint for a unified mass matrix model 
for quarks and leptons.

\vspace{10mm}

{\Large\bf Acknowledgment}   

The work is supported by JSPS 
(No.\ 21540266).

%


\end{document}